\documentclass[prb, aps, twocolumn, groupedaddress,showpacs]{revtex4}
\usepackage{eqnarray}
\usepackage[english]{babel} 
\usepackage[english]{layout}
\usepackage{amsmath,amsfonts,amssymb}
\usepackage{graphicx}
\usepackage{float}
\usepackage{color}
\usepackage{braket}
\usepackage{version}
\usepackage{array,multirow,makecell}

\begin{document}
\title{Measuring the Quantum Geometric Tensor in 2D Photonic and Polaritonic Systems}

\author{O.Bleu}
\author{D. D. Solnyshkov}
\author{G. Malpuech}
\affiliation{Institut Pascal, PHOTON-N2, Clermont Auvergne University, CNRS, 4 avenue Blaise Pascal, 63178 Aubi\`{e}re Cedex, France.} 

\begin{abstract}
We first consider a generic two-band model which can be mapped to a pseudospin on a Bloch sphere. We establish the link between the pseudospin orientation and the components of the quantum geometric tensor (QGT): the metric tensor and the Berry curvature. We show how the \textbf{k}-dependent pseudospin orientation can be measured in photonic systems with radiative modes. We consider the specific example: a 2D planar cavity with two polarization eigenmodes, where the pseudospin measurement can be performed via polarization-resolved photoluminescence. We also consider the s-band of a staggered honeycomb lattice for polarization-degenerate modes (scalar photons). The sublattice pseudospin can be measured by performing spatially resolved interferometric measurements.  In the second part, we consider a more complicated four-band model, which can be mapped to two entangled pseudospins. We show how the  QGT components can be obtained by measuring six angles. The specific four-band system we considered is the s-band of a honeycomb lattice for polarized (spinor) photons. We show that all six angles can indeed be measured in this system. We simulate realistic experimental situations in all cases. We find the photon eigenstates by solving Schr\"odinger equation including pumping and finite lifetime, and then simulate the measurement of the relevant angles to finally extract realistic mappings of the \textbf{k}-dependent QGT components.
\end{abstract}

\maketitle

\section{Introduction}

With the expansion of the field of topological physics, it is nowadays well understood that the knowledge of the spectrum of a Hamiltonian is not sufficient to have all the information of a quantum system. Indeed, the Berry curvature, determined by the eigenstates, is one of the central pillars of modern Physics \cite{berry1984quantal,RevModPhys.82.1959}. It plays the main role in a plethora of condensed matter phenomena. A local Berry curvature in momentum space affects the motion of particles and leads to intrinsic anomalous Hall and spin-Hall effects \cite{MacDonald2010,Sinova2015}. The integral of the Berry curvature over a complete band is a topological invariant which is associated with the existence of chiral conducting edge states as in  the quantum Hall effect, topological insulators and superconductors \cite{Hasan2010} and  also with the Fermi arc surface states in Weyl semi-metals \cite{armitage2017weyl}.

The Berry curvature is actually determined by the local geometry of quantum space, being included in a more general object -- the quantum geometric tensor (QGT).
This mathematical object was initially introduced in order to define the distance between quantum states \cite{provost1980riemannian}. It turns out that while its real part indeed defines a metric, its imaginary part is proportional to the Berry curvature \cite{berry1989quantum}.
The effects of the quantum metric on physical phenomena are less known than the ones of the Berry curvature, but there are several recent works highlighting the direct consequences of the quantum metric. In condensed matter, it appears to play a role in different contexts, ranging from orbital susceptibility \cite{Gao2015,Piechon2016} and corrections to the anomalous Hall effect \cite{Gao2014,bleu2016effective}, to the exciton Lamb shift in TMDs \cite{PhysRevLett.115.166802} and superfluidity in flat bands \cite{peotta2015superfluidity,PhysRevLett.117.045303}. Finally, the quantum metric is widely used to assess the fidelity in   quantum informatics \cite{mcmahon}.

Topological and Berry curvature-related single-particle phenomena have been extended from solid state physics to many other classical or quantum systems, such as photonic systems \cite{Onoda2004,Ozawa2014,lu2014topological}, where the analog of quantum Hall effect was first pointed out by Haldane and Raghu \cite{Haldane2008}, cold atoms \cite{Jotzu2014,goldman2016topological} and mechanical systems \cite{huber2016topological}, and more recently to geophysical waves \cite{delplace2017topological}.
The emulation of condensed matter Hamiltonians in artificial systems is an important part of modern Physics \cite{lewenstein2007ultracold,Bloch2008RevModPhys.80.885,Jotzu2014}. Recently, several protocols have been proposed to measure the Berry curvature in such systems \cite{Montambaux2015,Price2016,Hafezi2014} and some of them have been implemented experimentally \cite{flaschner2016experimental,wimmer2017experimental}. However, the real part of the QGT -- the quantum metric -- has never been measured experimentally, to our knowledge.  In a recent paper, T.~Ozawa proposes an experimental protocol to reconstruct the QGT components in a photonic flat band \cite{ozawa2017mapping}. This reconstruction is based on the anomalous Hall drift measurement of  the driven-dissipative stationary solution in different configurations, similar to previous works on the Berry curvature extraction \cite{Ozawa2014}.

In this paper, we propose a different method to extract the components of the quantum geometric tensor by \textit{direct} measurements using polarization-resolved and spatially resolved interference  techniques.
This proposal is based on the experimental ability to perform direct measurement of photon wave-function in radiative photonic systems such as planar cavities and cavity lattices \cite{Jacqmin2014,whittaker2017exciton}, but can be extended to other systems where \textbf{k}-dependent pseudospin orientations can be measured. 
Our method is designed to extract QGT components of systems with one or two coupled pseudospins (two-band or four-band models), independently of the band curvature. It can therefore be used for a wider range of systems than other recently proposed schemes, based on the anomalous Hall effect \cite{ozawa2017mapping}. 

We emphasize that our proposal concerns the measurement of geometrical quantities linked to the Hermitian part of the system. However, the dissipation (finite lifetime of the radiative states) is the key ingredient which enables the measurement. As highlighted in recent works, dissipation can also be linked to new topological numbers related to the non-hermiticity and the complex eigenenergies \cite{Leykam2017,shen2017topological}, but this is not the subject of the present work.

The outline of the paper is as follows. In section \ref{QGTdef} we quickly introduce the QGT.
Section \ref{twoband} is dedicated to two-band systems keeping in mind two particular implementations. The first case we consider is a planar microcavity taking into account the light polarization degree of freedom. The second case is a staggered honeycomb lattice (which can be made of coupled cavities) for scalar photons, where the pseudospin of interest is associated with the lattice degree of freedom. 
We generalize the measurement protocol to generic four-band systems described with two entangled pseudospins in section \ref{fourband}. This situation is realized in the s-band of a lattice with two atoms per unit cell (e.g. honeycomb lattice) taking into the polarization of light. It is also realized for scalar particles in the p-band of a honeycomb lattice. For all examples, in addition to the analytical and tight-binding results, we perform numerical simulations which aim to reproduce the experimental measurement. We solve numerically the Schr\"odinger equation including pumping and finite lifetime of the photonic states, we then extract the experimentally accessible parameters and use them to reconstruct the QGT components.

\section{Quantum geometric tensor}
\label{QGTdef}
We first introduce some useful mathematical definitions of the quantum geometric tensor.
This tensor can be defined in momentum parameter space as \cite{provost1980riemannian}:
\begin{equation}
{T_{ij}^n} = \left\langle {{\frac{\partial u_n}{{\partial {k_i}}} }}
 \mathrel{\left | {\vphantom {{\frac{\partial }{{\partial {k_i}}}\psi } {\frac{\partial }{{\partial {k_j}}}\psi}}}
 \right. \kern-\nulldelimiterspace}
 {{\frac{\partial u_n}{{\partial {k_j}}}}} \right\rangle  - \left\langle {{\frac{\partial u_n }{{\partial {k_i}}}}}
 \mathrel{\left | {\vphantom {{\frac{\partial }{{\partial {k_i}}}u_n} u_n}}
 \right. \kern-\nulldelimiterspace}
 {u_n} \right\rangle \left\langle {u_n}
 \mathrel{\left | {\vphantom {\psi  {\frac{\partial }{{\partial {k_j}}}\psi}}}
 \right. \kern-\nulldelimiterspace}
 {{\frac{\partial u_n}{{\partial {k_j}}}}} \right\rangle 
\end{equation}
An important property of this object is its gauge invariance, meaning that the components of the tensor are independent of the overall phase of the wavefunction $u_n$, where $n$ is the number of the band. Note that we use the notation $\ket{u_n}$ to describe quantum states even if not all the examples presented in the following are  based on periodic Bloch Hamiltonian. The parameter space of all Hamiltonians in this work is the reciprocal space $\mathbf{k}$. The real part of the QGT defines a metric, allowing to measure distances between the quantum states $\ket{u_n}$ in the $\mathbf{k}$-space, whereas its imaginary part defines the Berry curvature:
\begin{eqnarray}
g^n_{i,j}=\Re[T^n_{ij}]\\
B^n_l=-2\epsilon_{ijl}\Im[T^n_{ij}]
\end{eqnarray}
In the following, we consider two-dimensional systems ($i,j=x,y$), which means that $B_z$ is the only non-zero component of the Berry curvature.

The quantum metric and the Berry curvature can also be computed using the derivatives of the Hamiltonian instead of derivatives of the wavefunctions:
\begin{eqnarray}
g^n_{i,j}=\Re[\sum_{m\neq n}\frac{\bra{u_m}\partial_{i}H_k\ket{u_n}\bra{u_n}\partial_{j}H_k\ket{u_m}}{(E_m-E_n)^2}] \label{gHam} \\
B^n_l=i\epsilon_{ijl}\Im[\sum_{m\neq n}\frac{\bra{u_m}\partial_{i}H_k\ket{u_n}\bra{u_n}\partial_{j}H_k\ket{u_m}}{(E_m-E_n)^2}]\label{BHam}
\end{eqnarray}

However, this form is not convenient for direct experimental extraction, because only the wavefunction components can be measured and derived, and not the Hamiltonian itself.

\section{Two-band systems}
\label{twoband}
The Hamiltonian of any two-level (two-band) system can be mapped to a pseudospin coupled to an effective magnetic field, because the two-by-two Hamiltonian matrix can be decomposed into a linear combination of Pauli matrices and of the identity matrix. As shown below, the knowledge of the pseudospin is sufficient to reconstruct all the geometrical quantities linked with the eigenstates.
A general spinor wavefunction can be mapped on the Bloch sphere using two angles ($\theta$ - polar, $\phi$ - azimuthal) and written in circular polarization (spin-up, spin-down) basis:
\begin{equation}
\ket{u_{n,\mathbf{k}}}=\begin{pmatrix}\psi^+ \\ \psi^- \end{pmatrix}=\begin{pmatrix}\cos\frac{\theta\left(\mathbf{k}\right)}{2} e^{i\phi\left(\mathbf{k}\right)} \\ \sin\frac{\theta\left(\mathbf{k}\right)}{2} \end{pmatrix}
\label{eigvec1}
\end{equation}
with
\begin{eqnarray}
\theta\left(\mathbf{k}\right) =\arccos{S_z\left(\mathbf{k}\right)},~~ \phi=\arctan\frac{S_y\left(\mathbf{k}\right)}{S_x\left(\mathbf{k}\right)}
\label{anglez}
\end{eqnarray}
where the pseudospin components are linked with the intensity of each polarization of light, if the particular pseudospin is the Stokes vector of light:
\begin{eqnarray}
S_z =\frac{|\psi^+|^2-|\psi^-|^2}{|\psi^+|^2+|\psi^-|^2}\nonumber\\
S_x =\frac{|\psi^H|^2-|\psi^V|^2}{|\psi^H|^2+|\psi^V|^2} \label{pseudo}\\ 
S_y =\frac{|\psi^D|^2-|\psi^A|^2}{|\psi^D|^2+|\psi^A|^2}\nonumber
\end{eqnarray}
We remark here that pseudospin is arbitrary and can correspond to polarization pseudospin or to sublattice pseudospin if the system is a lattice with two atoms per unit cell. While for light the physical meaning of the vertical and diagonal polarizations is quite natural, for an arbitrary pseudospin they have to be reconstructed from the "circular" ($\psi_+$, $\psi_-$) basis as follows:
\[\begin{array}{l}
{\psi _H} = \frac{1}{{\sqrt 2 }}\left( {{\psi _ + } + {\psi _ - }} \right)\\
{\psi _V} = \frac{1}{{\sqrt 2 }}\left( {{\psi _ + } - {\psi _ - }} \right)\\
{\psi _D} = \frac{1}{{\sqrt 2 }}\left( {{e^{i\pi /4}}{\psi _ + } + {e^{ - i\pi /4}}{\psi _ - }} \right)\\
{\psi _A} = \frac{1}{{\sqrt 2 }}\left( {{e^{i\pi /4}}{\psi _ + } - {e^{ - i\pi /4}}{\psi _ - }} \right)
\end{array}\]

Applying Eq.~(1) to the eigenstates \eqref{eigvec1} leads to the formula:
\begin{eqnarray}
g_{ij} =\frac{1}{4}(\partial_i\theta\partial_j\theta+\sin^2\theta \partial_i\phi\partial_j\phi)\\
B =\frac{1}{2}\sin\theta(\partial_x\theta\partial_y\phi-\partial_y\theta\partial_x\phi)
\label{xtract}
\end{eqnarray}
where $i$,$j$ indices stand for $k_x$,$k_y$ components.
Therefore, extracting $\theta(\mathbf{k})$ and $\phi(\mathbf{k})$ for a given energy band at each wavevector $\mathbf{k}$ allows to fully reconstruct the components of the QGT in momentum space. This protocol can be implemented using polarization-resolved photoluminescence or interferometry techniques available for light in the state-of-the-art experiments.  For two-band systems, the metric tensor is the same for each band ($g_{ij}^+=g_{ij}^-=g_{ij}$), whereas the Berry curvatures are opposite ($B^+=-B^-$) \cite{Piechon2016}.

\subsection{Planar cavity}
A planar microcavity has two main features important for our study. First, it has a two-dimensional parabolic dispersion of photons close to zero in-plane wavevector, because of the quantization in the growth direction. This allows to use the Schr\"odinger formalism to deal with massive photons. Second, the energy splitting between TE and TM polarized eigen modes is analogous to a spin-orbit coupling for photons \cite{Kavokin2005}, which 
is a necessary ingredient to obtain a non-zero Berry curvature. The other necessary ingredient to get non-zero Berry curvature is an effective Zeeman splitting, which in practice can be implemented by using strong coupling of cavity photons and quantum well excitons, achieved in modern microcavities \cite{Weisbuch1992}. The excitons are sensitive to applied magnetic fields: they exhibit a Zeeman splitting between the components coupled to $\sigma^+$ and $\sigma^-$-polarized photons, inducing a Zeeman splitting for the resulting quasiparticles - exciton-polaritons \cite{Magnetop2008}.

\begin{figure}[tbp]
 \begin{center}
 \includegraphics[scale=0.4]{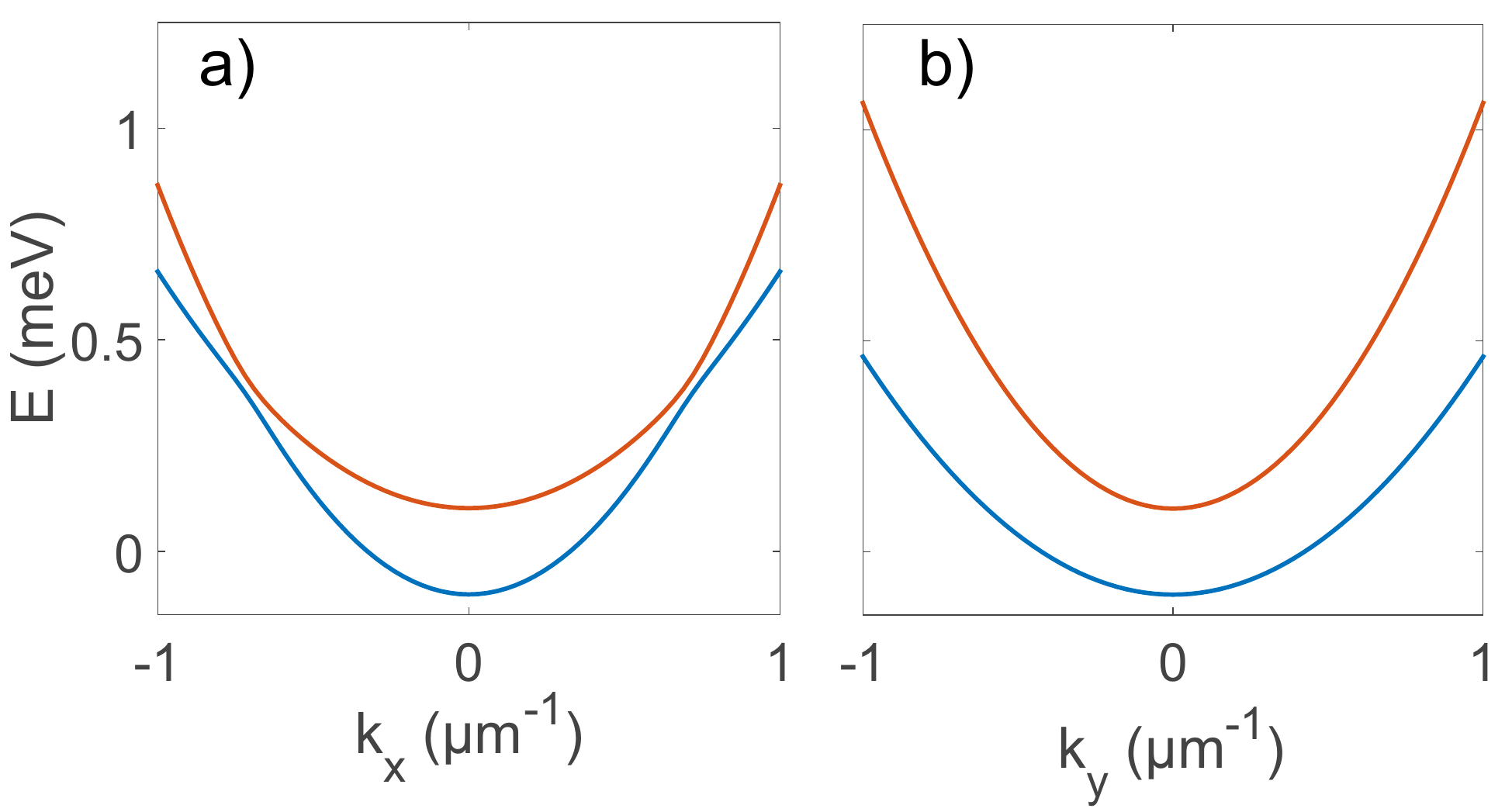}
 \caption{Dispersion of the planar microcavity with crossed effective magnetic fields ($\Omega_x$ and $\Omega_z$ and TE-TM splitting. a) $k_x$ cross-section, b) $k_y$ cross-section.}
 \label{Fig1}
  \end{center}
 \end{figure}

Here, we consider an additional splitting between linear polarizations which acts as a static in-plane field \cite{martin2006optical}. Such field, usually linked with the cristallographic axes, can appear because of the anisotropy of the quantum well, and it can be controlled by an electric field applied in the growth direction \cite{glazov2006}.
The resulting Hamiltonian in momentum space can be written as a two-by-two matrix in circular basis $(\psi^+,\psi^-)^T$. 

\begin{eqnarray}
H_{k}=\begin{pmatrix}
\frac{\hbar^2k^2}{2m^*}+\Delta_z &\alpha e^{-i\varphi_0}+\beta k^2 e^{2i\varphi} \\
\alpha e^{i\varphi_0}+\beta k^2 e^{-2i\varphi}& \frac{\hbar^2k^2}{2m^*}-\Delta_z
\end{pmatrix} 
\end{eqnarray}
where $\alpha$, $\beta$, and $\Delta_z$ define the strength of the effective fields corresponding to the constant X-Y splitting, TE-TM spin-orbit coupling, and the Zeeman splitting, respectively. $m^*=m_lm_t/(m_l+m_t)$, with $m_l$ and $m_t$ corresponding to the longitudinal and transverse effective masses. $k=\sqrt{k_x^2+k_y^2}$ is the in-plane wavevector with $k_x=k\cos{\varphi}$, $k_y=k\sin{\varphi}$. $\varphi_0$ is the in-plane angle of the constant field. The eigenvalues of this Hamiltonian for realistic parameters are shown in Fig.~\ref{Fig1} as the cross-sections of the 2D dispersion in the $k_x$ and $k_y$ directions.

\begin{figure}[tbp]
 \begin{center}
 \includegraphics[scale=0.4]{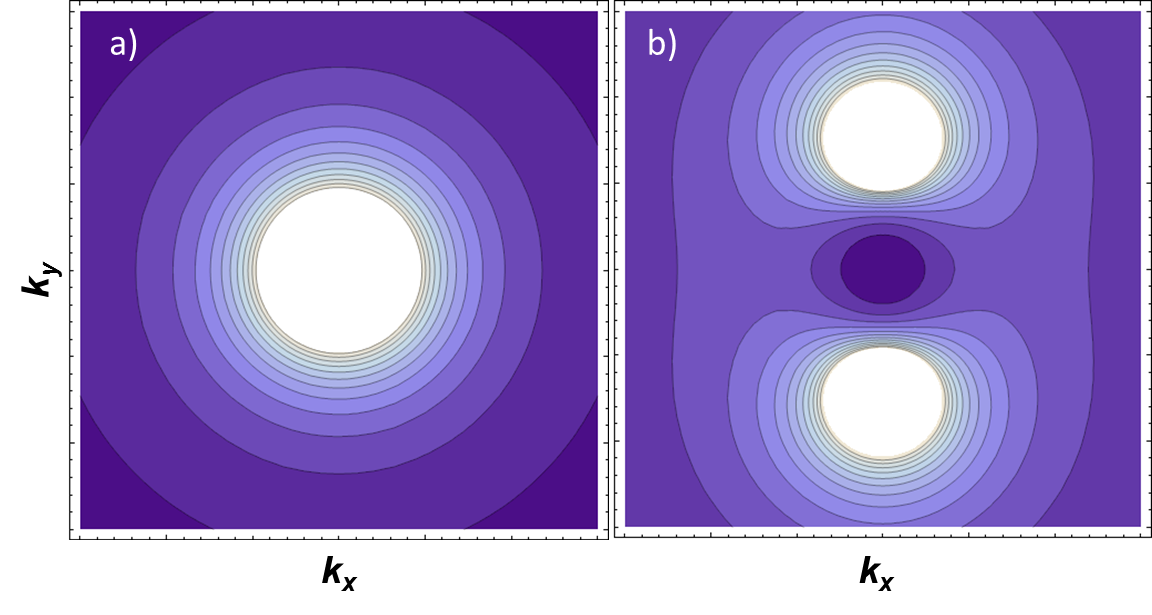}
 \caption{Trace of the metric tensor $g_{xx}+g_{yy}$ of the LPB in a cavity system: without (a) and with XY splitting (b) from the analytical formula \eqref{xyQGT}.}
 \label{Fig2b}
  \end{center}
 \end{figure}

Choosing $\varphi_0=0$, which means that the constant field is in the $x$ direction, the QGT components are found analytically:
\begin{eqnarray}
g_{xx}=\frac{\beta^2\left(k_y^2\left(\alpha- k^2\beta \right)^2+k^2\Delta_z^2 \right)}{\left(\alpha^2+2\left(k_x^2-k_y^2 \right)\alpha\beta+k^4\beta^2+\Delta_z^2 \right)^2}\nonumber \\
g_{yy}=\frac{\beta^2\left(k_x^2\left(\alpha+ k^2\beta \right)^2+k^2\Delta_z^2 \right)}{\left(\alpha^2+2\left(k_x^2-k_y^2 \right)\alpha\beta+k^4\beta^2+\Delta_z^2 \right)^2} \nonumber \\
g_{xy}=\frac{\beta^2k_xk_y\left(\alpha^2- k^4\beta^2 \right)}{\left(\alpha^2+2\left(k_x^2-k_y^2 \right)\alpha\beta+k^4\beta^2+\Delta_z^2 \right)^2} \nonumber \\
B_{\pm}=\frac{\pm2\beta^2k^2\Delta_z}{\left(\alpha^2+2\left(k_x^2-k_y^2 \right)\alpha\beta+k^4\beta^2+\Delta_z^2 \right)^{3/2}} 
\label{xyQGT}
\end{eqnarray}
We see that while the Berry curvature requires a non-vanishing Zeeman splitting, the metric tensor can be nonzero even without any applied magnetic field: the TE-TM spin-orbit coupling is sufficient. We plot the calculated trace of the metric tensor as a function of wavevector for $\beta=0.1$ in the absence of Zeeman splitting ($\Delta_z=0$) in Fig. \ref{Fig2b}. Panel (a) exhibits cylindrical symmetry due to $\alpha=0$, while panel (b) demonstrates the transformation of the metric in the reciprocal space in presence of non-zero in-plane effective field $\alpha=0.2$. We stress that the metric diverges where the states become degenerate (an emergent non-Abelian gauge field forms around these points \cite{Tercas2014} when $\alpha\neq 0$), but it can nevertheless be measured sufficiently far from the points of degeneracy.

\begin{figure}[tbp]
 \begin{center}
 \includegraphics[scale=0.4]{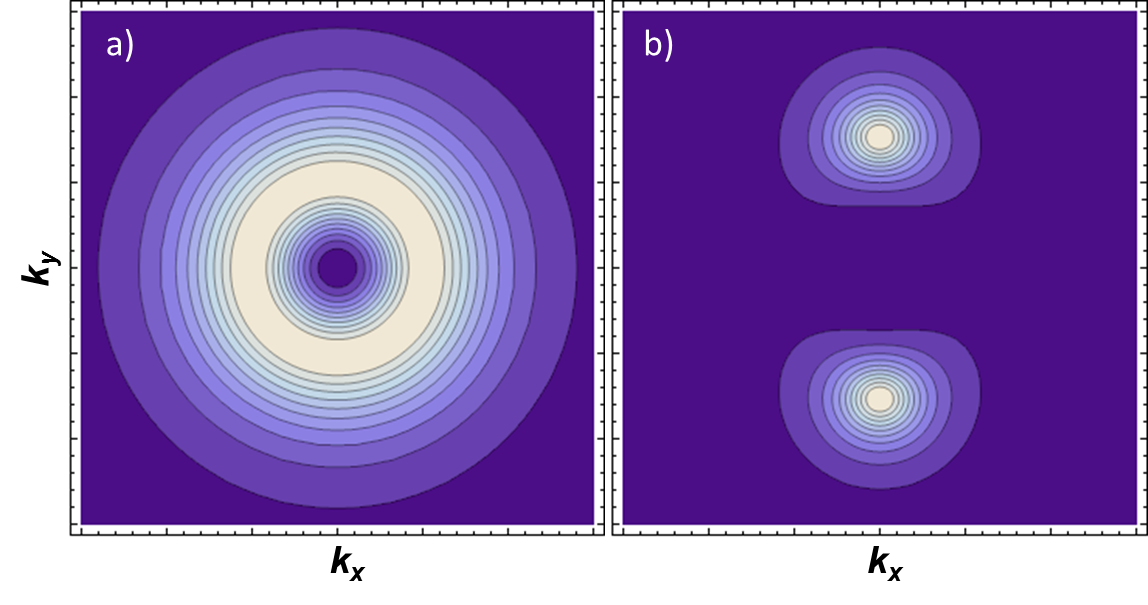}
 \caption{Berry curvature of the LPB in a cavity system: without (a) and with XY splitting (b) from the analytical formula \eqref{xyQGT}.}
 \label{Fig2}
  \end{center}
 \end{figure}

Next, we plot the Berry curvature for a non-zero Zeeman splitting $\Delta_z=0.1$ in Fig. \ref{Fig2}. Note, that a $\alpha\neq0$ implying anisotropic eigenenergies  leads to an important change in the Berry curvature distribution in momentum space from a ring-like maximum to two point-like maxima in the $k_y$ direction, similar to what happened to the metric tensor. Actually, Berry curvature is highest at the anticrossing of the branches, where the metric tensor was divergent for zero Zeeman splitting. In the isotropic case, this anticrossing does not depend on the direction of the wavevector, while the in-plane field breaks this isotropy and gives two preferential directions for the anticrossing, where the TE-TM splitting and the in-plane field compensate each other (see Fig.~\ref{Fig1}).\\

These results can be directly compared with numerical simulations, from which the QGT components are extracted using Eq. \eqref{xtract}. Here, and in the following, we are solving the 2D Scr\"odinger equation numerically over time:
\begin{eqnarray}\label{schro}
& i\hbar \frac{{\partial \psi _ \pm  }}
{{\partial t}}  =  - \frac{{\hbar ^2 }}
{{2m}}\Delta \psi _ \pm  - \frac{{i\hbar }}
{{2\tau }}\psi _ \pm  \pm\Delta_z\psi_\pm\\
& + \beta {\left( {\frac{\partial }{{\partial x}} \mp i\frac{\partial }{{\partial y}}} \right)^2}{\psi _ \mp } +\alpha e^{\mp i\varphi_0}\psi_\mp +U\psi_\pm + \hat{P} \nonumber 
\end{eqnarray}
where ${\psi_+(\mathbf{r},t), \psi_-(\mathbf{r},t)}$ are the two circular components, $m=5\times10^{-5}m_{el}$ is the polariton mass, $\tau=30$ ps the lifetime, $\beta$ is the TE-TM coupling constant (corresponding to a 5\% difference in the longitudinal and transverse masses). $\Delta_z=0.06$ meV is the magnetic field in the $Z$ direction (Zeeman splitting), $\alpha$ is the in-plane effective magnetic field (splitting between linear polarizations) with its orientation given by $\varphi_0=0$, $\hat{P}$ is the pump operator (Gaussian noise or Gaussian pulse exciting all states at $t=0$). $U$ is an external potential used in the following sections to encode the lattice potential (here, $U=0$). 

\begin{figure}[tbp]
 \begin{center}
 \includegraphics[scale=0.55]{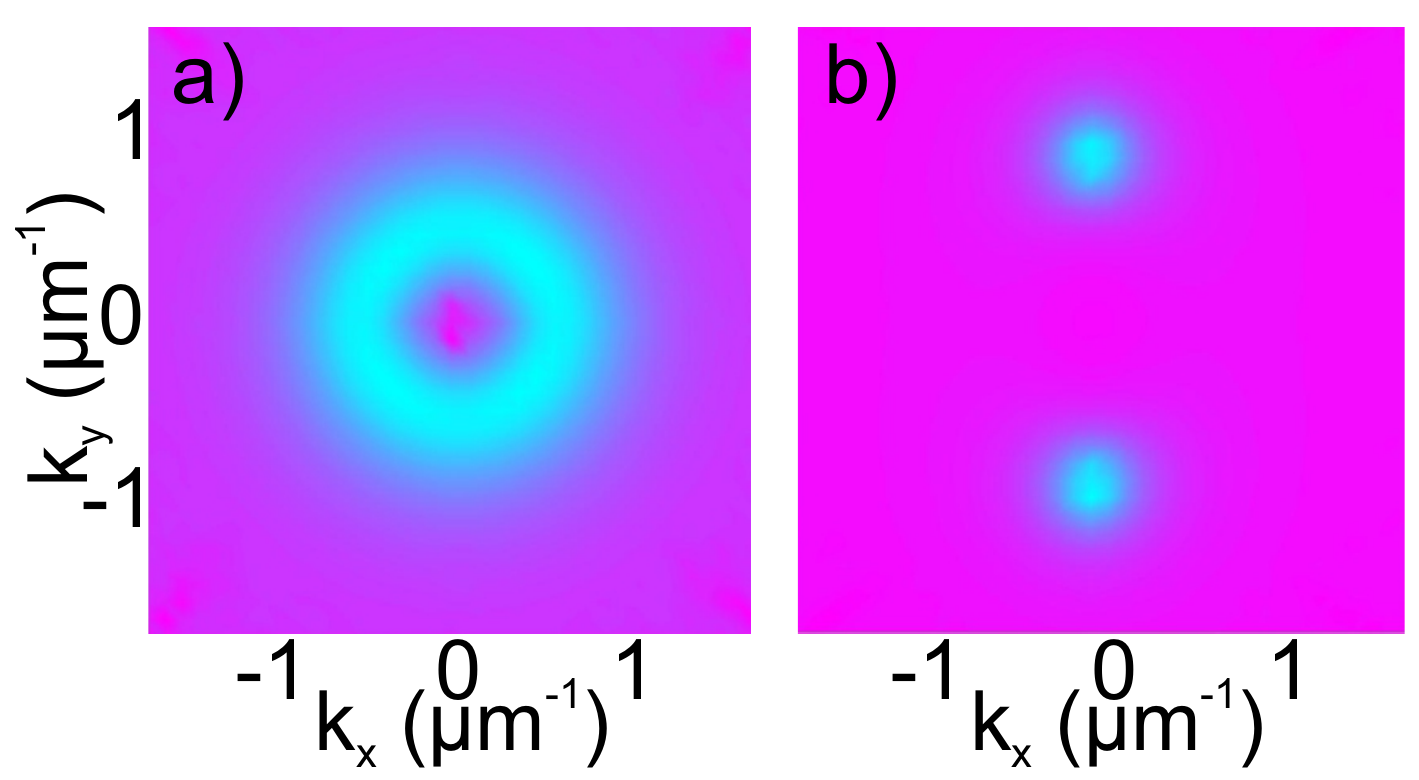}
 \caption{Berry curvature extracted from numerical simulations using Schr\"odinger equation and eq. \eqref{xtract}.}
 \label{Fig3}
  \end{center}
 \end{figure}

The solution of this equation is then Fourier-transformed $\psi(\mathbf{r},t)\to\psi(\mathbf{k},\omega)$ and analyzed as follows. For each wavevector $\mathbf{k}$, we find the corresponding eigenenergy as a maximum of $|\psi(\mathbf{k},\omega)|^2$ over $\omega$. Then, the pseudospin $\mathbf{S}$ and its polar and azimuthal angles $\theta,\phi$ are calculated from the wavefunction $\psi(\mathbf{k},\omega)$ using Eqs. \eqref{anglez}-\eqref{pseudo}. This corresponds to optical measurements of all 6 polarization projections at a given wavevector and energy. Finally, the Berry curvature is extracted from $\theta(\mathbf{k}),\phi(\mathbf{k})$ using Eq. \eqref{xtract}. The results are shown in Fig.~\ref{Fig3}. Panel (a) shows the Berry curvature in a planar cavity without the in-plane splitting ($\alpha=0$). Panel (b) demonstrates the modification of the Berry curvature under the effect of a non-zero in-plane field $\alpha=0.1$~meV. As in the analytical solution, the ring is continuously transformed into two maxima.

\subsection{Staggered honeycomb lattice for scalar particles}
The Hamiltonian of a staggered honeycomb lattice for scalar particles, in the tight-binding approximation with two atoms per unit cell, is also a two-by-two matrix which can be mapped to an effective magnetic field acting on the \textit{sublattice} pseudospin. 
The Bloch Hamiltonian in $(\psi_A,\psi_B)^T$ basis reads \cite{Wallace1947}:
\begin{eqnarray}
H_{k}=\begin{pmatrix}
\Delta_{AB} &-J f_k \\
-J f_k^*& -\Delta_{AB}
\end{pmatrix} 
\end{eqnarray}
where  $f_k=\sum_{j=1}^3\exp{(-i\textbf{kd}_{\phi_j})}$ and $\Delta_{AB}$ is energy difference between A and B sublattice states.
 \begin{figure}[tbp]
 \begin{center}
 \includegraphics[scale=0.45]{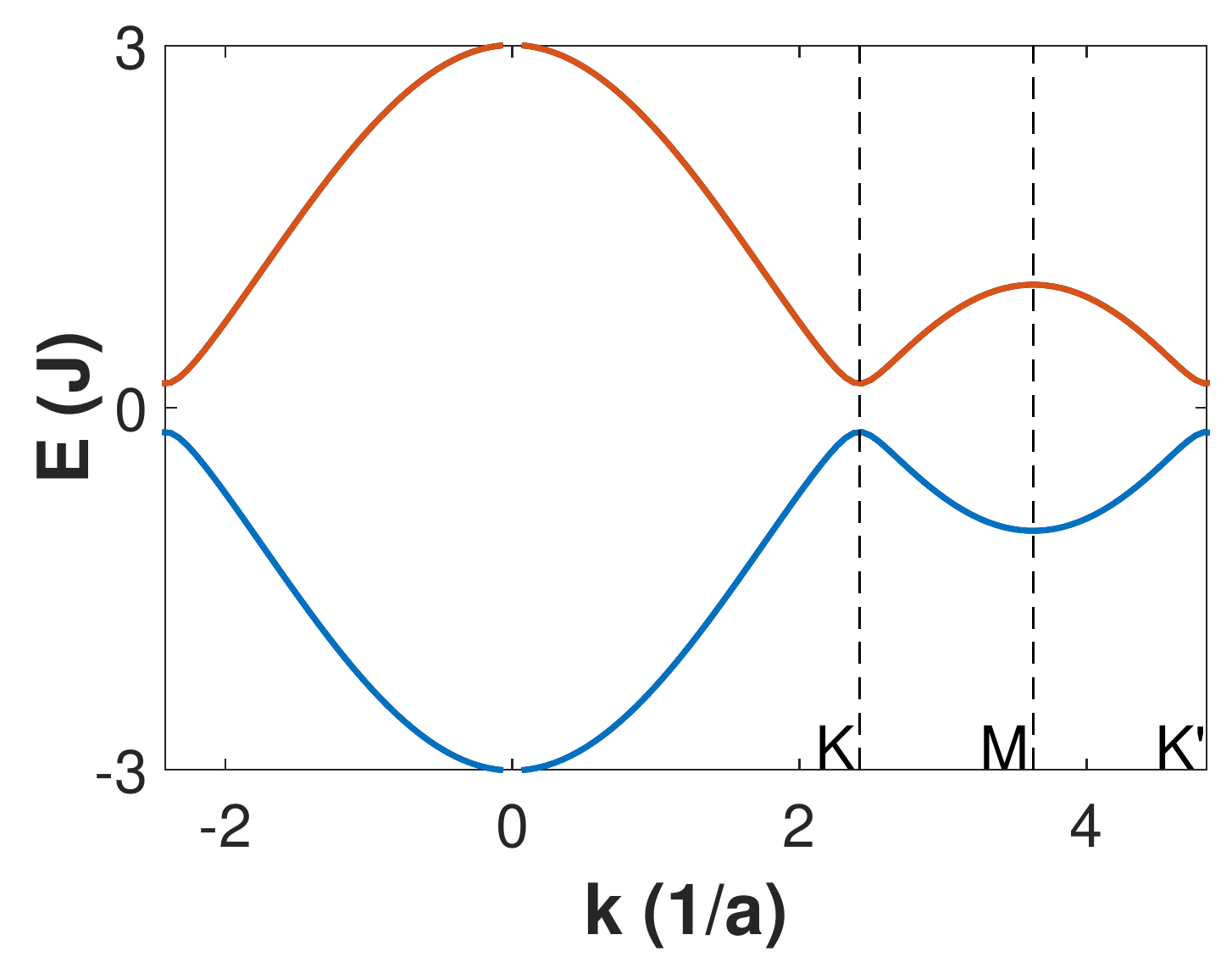}
 \caption{Staggered honeycomb lattice tight-binding dispersion ($\Delta_{AB}/J=0.2$). Dashed vertical lines mark high symmetry points in the first Brillouin zone.}
 \label{Fig4}
  \end{center}
 \end{figure}
The corresponding tight-binding dispersion is plotted in Fig.~\ref{Fig4}.
The gap, opened by the staggering potential, leads to opposite Berry curvatures at $K$ and $K'$ points \cite{Niu2007,RevModPhys.82.1959}. While simple analytical formula can be achieved by linearization of the Hamiltonian around these points, here, we compute the geometrical quantities numerically using eqs. \eqref{gHam}, \eqref{BHam} which, thanks to a better precision, allows to recover the signature of the underlying lattice in the QGT components (Fig.~\ref{Fig5}). Indeed, the presence of two valleys in the hexagonal Brillouin zone  implies a triangular shape of QGT components around $K$ and $K'$ points, which is neglected in the first-order approximation.\\

 \begin{figure}[tbp]
 \begin{center}
 \includegraphics[scale=0.65]{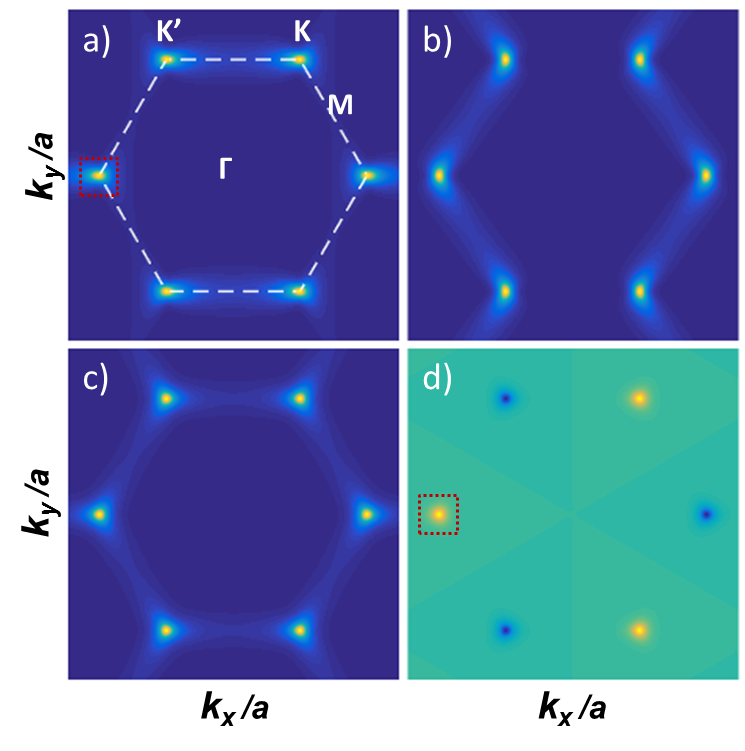}
 \caption{Quantum geometric tensor components in staggered honeycomb lattice (tight-binding results). (a) $g_{xx}$, (b) $g_{yy}$, (c) $g_{xx}+g_{yy}$, (d) Berry curvature $B_z$. (lower band, $\Delta_{AB}/J=0.2$). Dashed red squares around $K$ point show the zoomed region for the numeric QGT extraction.}
 \label{Fig5}
  \end{center}
 \end{figure}
 
 \begin{figure}[tbp]
 \begin{center}
 \includegraphics[scale=0.6]{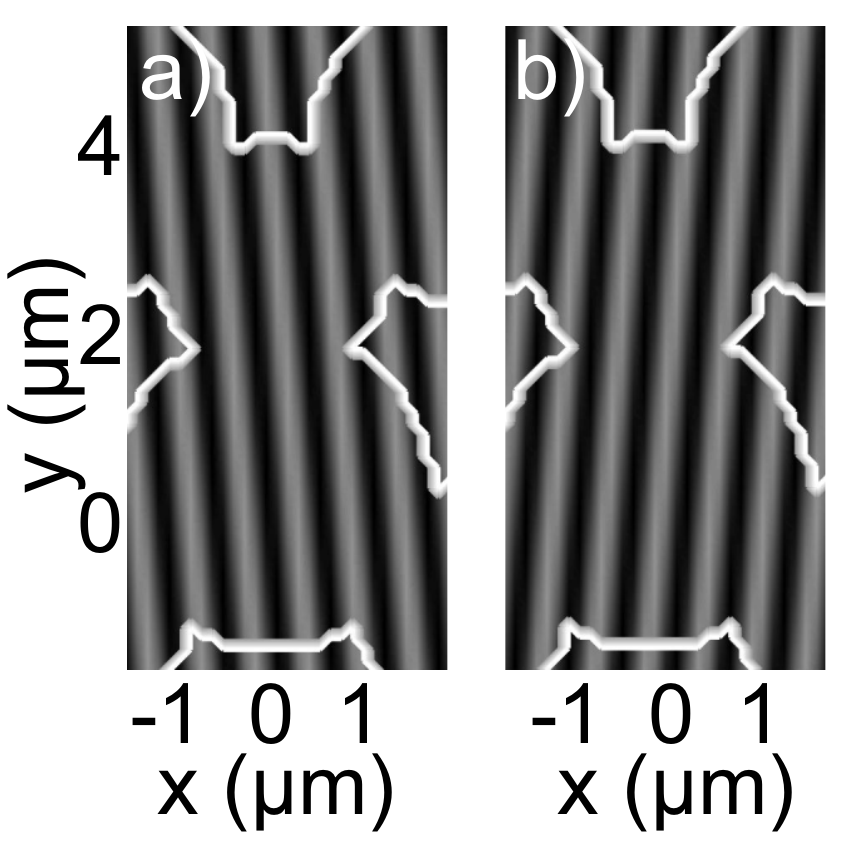}
 \caption{Examples of interference pattern in real space used in order to reconstruct phase difference between A and B pillars for two opposite values of $q$.}
 \label{Fig7}
  \end{center}
 \end{figure}
 
 \begin{figure}[tbp]
 \begin{center}
 \includegraphics[scale=0.55]{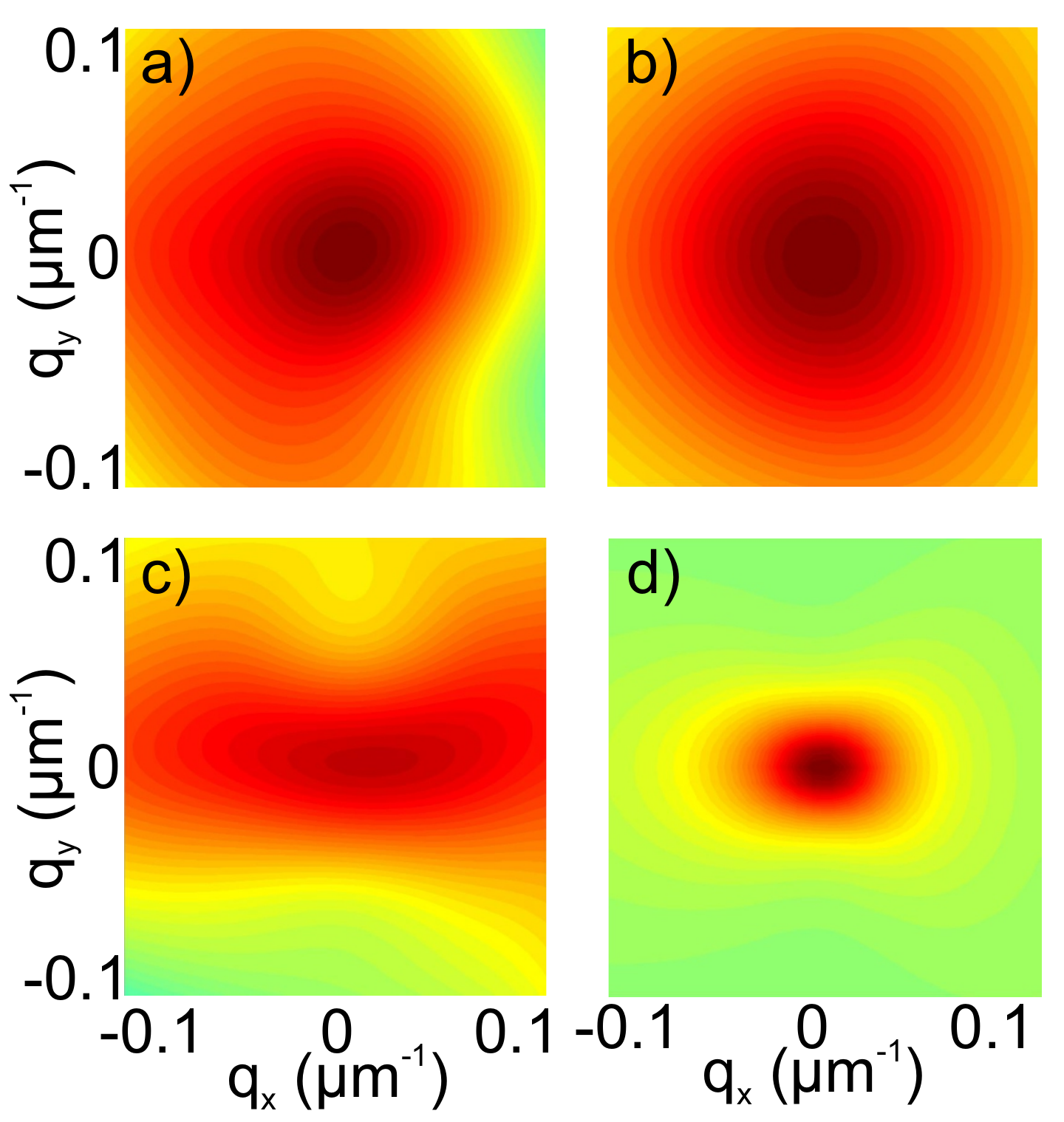}
  \caption{Berry curvature (a) and metric tensor component $g_{xx}$ (c),  extracted from numerical simulations based on the Schr\"odinger equation \eqref{schro}, compared with corresponding tight-binding results (b,d). Zoom around $K$ point (dashed red square in Fig.~\ref{Fig5}).}
 \label{Fig6}
  \end{center}
 \end{figure}

We have also performed numerical simulations with the QGT extraction for the staggered honeycomb lattice. In this section, to consider scalar particles, \textbf{only one spin component} was taken into account in the Schr\"odinger equation \eqref{schro} ($\psi_+$) and all coupling between the components was removed ($\alpha=0$, $\beta=0$). Thus, the only remaining pseudospin is the sublattice pseudospin linked with the honeycomb potential encoded in $U(\textbf{r})\neq 0$. We use a lattice potential $U(\textbf{r})$ of $26\times 26$ unit cells with radius of the pillars $r=1.5\mu m$, pillar radius modulation of 30\%,  and lattice parameter $a=2.5\mu m$. 

Once the wavefunction $\psi(\mathbf{r},t)$ and its image $\psi(\mathbf{k},\omega)$ are found, we extract the angles $\theta$ and $\phi$ defining the spinor. The physical meaning of the spinor here is different from that of the previous section, and the meaning of these angles differs as well.  For $S_z$ and $\theta$ the measurement is straightforward, because $|\psi_+|^2$ and $|\psi_-|^2$ are simply the intensities of emission from the two pillars $A$ and $B$ in the unit cell. To determine $\phi$, the phase difference between the two pillars, one has to consider the real space Fourier image of the corresponding wavevector state (the Bloch wave in real space) and determine this phase by interference measurements with a reference beam. This technique is analogous to the one used recently to measure the phase difference between pillars in a honeycomb photonic molecule \cite{Sala2015}. Figure~\ref{Fig7} shows two interference patterns for two opposite wavevectors $q$ close to a particular Dirac point $K$. The reference beam propagates along the $x$ direction, and the deviation of the interference fringes from the vertical direction is an evidence for the phase difference between the pillars.

Figure \ref{Fig6} shows the results of the extraction of the QGT components as discussed above. Panel (a) shows the Berry curvature $B_Z$, and panel (c) shows the $XX$ component of the quantum metric ($g_{xx}$), with corresponding tight-binding results shown in panels (b) and (d), respectively. All panels are shown in the vicinity of one of the Dirac points (chosen as the reference for the wavevector $\mathbf{q}$), where these components differ from zero. This allows to demonstrate that the resolution of the method is sufficient for the extraction in spite of the broadening due to the finite lifetime, numerical disorder, and the finite size of the structure. We see that the $g_{xx}$ component is compressed along the vertical direction, as in the tight-binding calculation (Fig.~\ref{Fig5}(a), and zoom in Fig. \ref{Fig6}(d)), and that the Berry curvature shows a slight triangular distortion due to the symmetry of the valley \cite{Nalitov2015b}, which would be simply cylindrical in the first order.

\section{four-band systems}
\label{fourband}
Several systems are well described by four-band Hamiltonians. Some examples are bilayer honeycomb lattices \cite{mccann2013electronic}, spinor s-bands  or p-bands in lattices with two atoms per unit cell \cite{Wu2008}.

When it comes to accounting for an additional degree of freedom like polarization pseudospin in a two-band lattice system, where there is already a sublattice pseudospin, one may think that measuring the two pseudospins should be sufficient to deduce the QGT in the first Brillouin zone.

It is indeed the case when the Hamiltonian can be decomposed in two uncoupled two-by-two blocks, which means that the two pseudospins are independent. This situation is realized for fermions in lattices in presence of time reversal symmetry for instance \cite{Kane2005,Bernevig2006}, where the two pseudospins are Kramers partners. Here, we consider a more generic situation, where we account for the possible coupling of the two pseudospins: an eigenstate of the full system cannot be decomposed as a product of the two pseudospins. The wavefunction has to take into account the entanglement of the two subsystems. A general 4-component wavefunction can be written as (see Appendix for an extended discussion of the generality):
\begin{eqnarray}
\ket{u_{n,\mathbf{k}}}&=&\left(c_A^+ e^{i\phi_A^+},c_A^-e^{i\phi_A^-},c_B^+e^{i\phi_B^+},c_B^-e^{i\phi_B^-}\right)^T \nonumber\\
&=&e^{i\phi_B^-}  \begin{pmatrix}
\cos\frac{\theta^A}{2}\cos\frac{\theta^{AB}}{2} e^{i\phi_A} e^{i\phi_{AB}}\\\sin\frac{\theta^A}{2}\cos\frac{\theta^{AB}}{2}e^{i\phi_{AB}}\\ \cos\frac{\theta^B}{2}\sin\frac{\theta^{AB}}{2} e^{i\phi_{B}}\\\sin\frac{\theta^B}{2}\sin\frac{\theta^{AB}}{2}\label{eigvec2}
\end{pmatrix}
\end{eqnarray}
Hence, six angles are necessary to parametrize the general wavefunction. As in the previous section, they are related to pseudospin components:
\begin{eqnarray}
\phi_A=\phi_A^+-\phi_A^-=\arctan\frac{S_y^A}{S_x^A}\nonumber\\
\phi_B=\phi_B^+-\phi_B^-=\arctan\frac{S_y^B}{S_x^B}\\
\phi_{AB}=\phi_A^--\phi_B^-=\arctan\frac{S_y^{AB^-}}{S_x^{AB^-}}\nonumber
\end{eqnarray}
and
\begin{eqnarray}
\theta_A=\arccos{S_z^{A}}\nonumber\\
\theta_B=\arccos{S_z^{B}}\\
\theta_{AB}=\arccos{S_z^{AB}}\nonumber
\end{eqnarray}
where $\phi_A$, $\phi_B$, $\theta_A$, $\theta_B$ are defined by the internal pseudospin (eg. polarization) on each component of the external pseudospin (A/B sublattices), $\phi_{AB}$ is the phase difference between the sublattice components for a given component ($\sigma^-$) of the internal pseudospin. $\theta_{AB}$ is defined by the \emph{total} intensity difference between the two sublattices. The measurement of these six angles in a band allows a full reconstruction of the corresponding eigenstate. 
Using the eigenstate formulation \eqref{eigvec2}, one can derive the QGT component formulas in terms of these angles:

\begin{widetext}
\begin{eqnarray}
g_{ij}=&&\frac{1}{4}(\partial_i\theta^{AB}\partial_j\theta^{AB}+\partial_i\theta^{A}\partial_j\theta^{A}\cos^2\frac{\theta^{AB}}{2}+\partial_i\theta^{B}\partial_j\theta^{B}\sin^2\frac{\theta^{AB}}{2}+\partial_i\phi^{AB}\partial_j\phi^{AB}\sin^2\theta^{AB} \nonumber\\&&+\cos^2\frac{\theta^A}{2}\cos^2\frac{\theta^{AB}}{2}(3-\cos\theta^{AB}-\cos\theta^{A}(1+\cos\theta^{AB}))\partial_i\phi^{A}\partial_j\phi^{A}\nonumber\\&&+\cos^2\frac{\theta^B}{2}\sin^2\frac{\theta^{AB}}{2}(3+\cos\theta^{AB}+\cos\theta^{B}(\cos\theta^{AB}-1))\partial_i\phi^{B}\partial_j\phi^{B}\nonumber\\&&+\cos^2\frac{\theta^A}{2}\sin^2\theta^{AB}(\partial_i\phi^{AB}\partial_j\phi^{A}+\partial_j\phi^{AB}\partial_i\phi^{A})\nonumber\\&&-\cos^2\frac{\theta^B}{2}\sin^2\theta^{AB}(\partial_i\phi^{AB}\partial_j\phi^{B}+\partial_j\phi^{AB}\partial_i\phi^{B})\nonumber\\&&-\cos^2\frac{\theta^A}{2}\cos^2\frac{\theta^B}{2}\sin^2\theta^{AB}(\partial_i\phi^{A}\partial_j\phi^{B}+\partial_j\phi^{A}\partial_i\phi^{B}))
\label{gextract}
\end{eqnarray}

\begin{eqnarray}
B_{z} =&&\frac{1}{4}(\sin\theta^A\cos^2\frac{\theta^{AB}}{2}(\partial_x\theta^A\partial_y\phi^A-\partial_y\theta^A\partial_x\phi^A)+\sin\theta^B\sin^2\frac{\theta^{AB}}{2}(\partial_x\theta^B\partial_y\phi^B-\partial_y\theta^B\partial_x\phi^B)\nonumber \\&&+\sin\theta^{AB}\cos^2\frac{\theta^{A}}{2}(\partial_x\theta^{AB}\partial_y\phi^A-\partial_y\theta^{AB}\partial_x\phi^A)-\sin\theta^{AB}\cos^2\frac{\theta^{B}}{2}(\partial_x\theta^{AB}\partial_y\phi^B-\partial_y\theta^{AB}\partial_x\phi^B)\nonumber \\&&+\sin\theta^{AB}(\partial_x\theta^{AB}\partial_y\phi^{AB}-\partial_y\theta^{AB}\partial_x\phi^{AB}))
\label{Bextract}
\end{eqnarray}
\end{widetext}
One can observe that the formula complexity has clearly increased compared to the two-state system. However, we stress that if the energy spectrum is accessible experimentally with sufficient resolution, the extraction protocol difficulty does not increase despite the higher number of angles. 
 In the following, we use a specific case in order to demonstrate the feasibility of the measurement.

\subsection*{Honeycomb lattice for spinor particles}

In this section, we consider the s-band of a regular honeycomb lattice containing vectorial (polarized) photons with TE-TM splitting and an external Zeeman field as an example of a four-state system. 
In such system, the quantum anomalous Hall effect for polaritons has been predicted recently \cite{Nalitov2015}.
The minimal tight-binding Bloch Hamiltonian written in circular basis $(\psi_A^+,\psi_A^-,\psi_B^+,\psi_B^-)^T$ is the following:
\begin{eqnarray}
H_k=\begin{pmatrix}
\Delta_z\sigma_z&F_k \\
F_k^+&\Delta_z\sigma_z
\end{pmatrix} 
,\quad F_k=-\begin{pmatrix}
f_kJ&f_k^+\delta J \\
f_k^- \delta J&f_kJ
\end{pmatrix}
\end{eqnarray}
where $\delta J$ is the TE-TM SOC strength and $f_k^{\pm}=\sum_{j=1}^3 \exp{(-i[\textbf{kd}_{\phi_j}\mp 2\phi_j])}$. $\Delta_z$ is the Zeeman field and $\sigma_z$ the third Pauli matrix. 
 \begin{figure}[tbp]
 \begin{center}
 \includegraphics[scale=0.45]{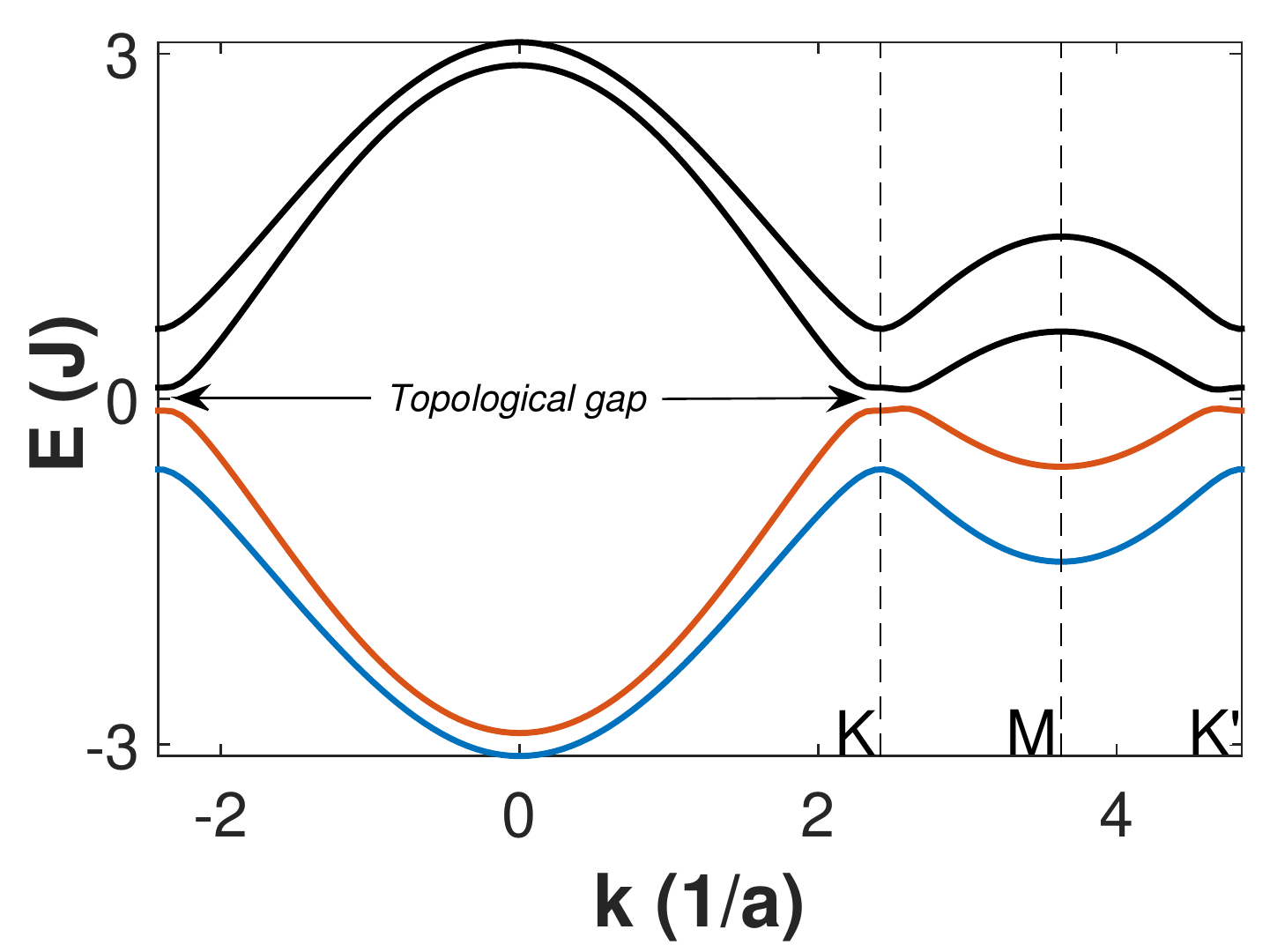}
 \caption{Tight-binding dispersion of regular honeycomb lattice with TE-TM SOC and Zeeman field ($\Delta_z/J=0.1$, $\delta J/J=0.2$). Dashed vertical lines mark high symmetry points in the first Brillouin zone.}
 \label{Fig8}
  \end{center}
 \end{figure}

  \begin{figure}[tbp]
 \begin{center}
 \includegraphics[scale=0.6]{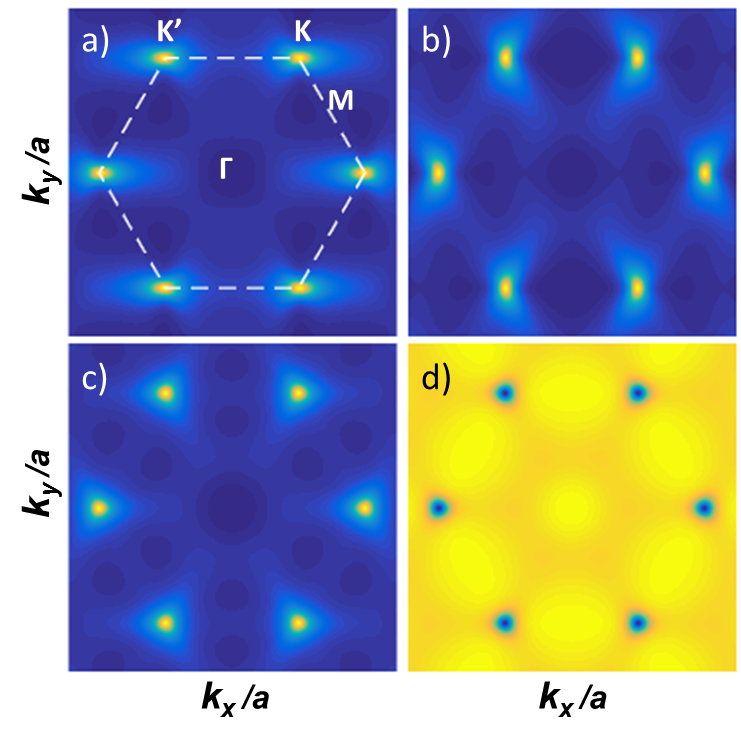}
 \caption{Quantum geometric tensor components in regular honeycomb lattice (1st band, tight-binding results). (a) $g_{xx}^1$, (b) $g_{yy}^1$, (c) $g_{xx}^1+g_{yy}^1$, (d) Berry curvature $B_z^1$. (1st band, $\Delta_z/J=0.1$, $\delta J/J=0.2$) \label{Fig9}}
  \end{center}
 \end{figure} 
 
The Hamiltonian becomes four-by-four matrix due to the additional polarization degree of freedom. The typical dispersion in the first Brillouin zone is plotted on Figure \ref{Fig8}. This time the full bandgap between the two lower and two upper bands is opened thanks to the combination of the Zeeman field (which breaks time-reversal symmetry) and the TE-TM SOC. In this configuration, the Berry curvatures around $K$ and $K'$ point have the same sign and the Chern number characterizing the bandgap is non-zero.

  \begin{figure}[tbp]
 \begin{center}
 \includegraphics[scale=0.6]{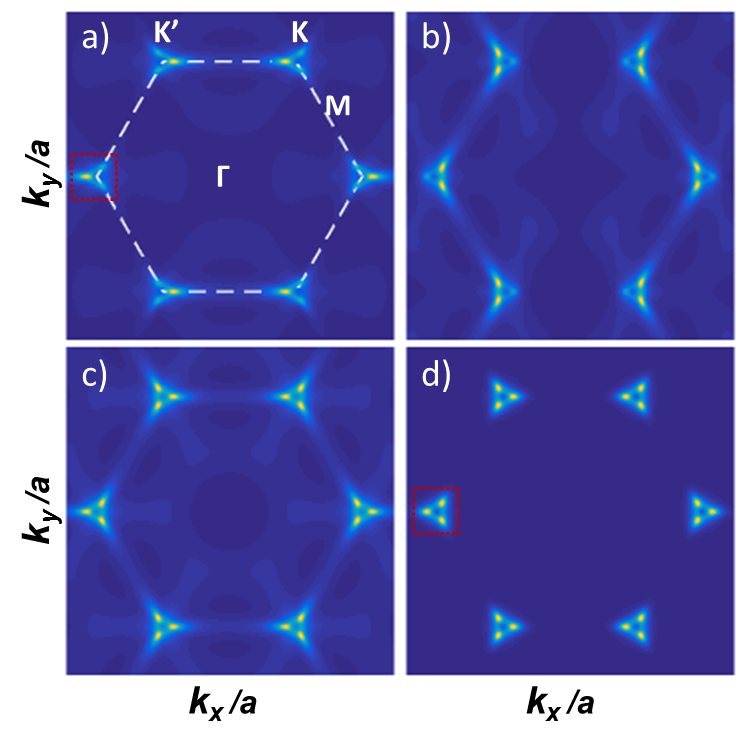}
 \caption{Quantum geometric tensor components in regular honeycomb lattice (tight-binding results). (a) $g_{xx}^2$, (b) $g_{yy}^2$, (c) $g_{xx}^2+g_{yy}^2$, (d) Berry curvature $B_z^2$. (2nd band, $\Delta_z/J=0.1$, $\delta J/J=0.2$) \label{Fig10}}
  \end{center}
 \end{figure}
\begin{figure}[h]
 \begin{center}
 \includegraphics[scale=0.55]{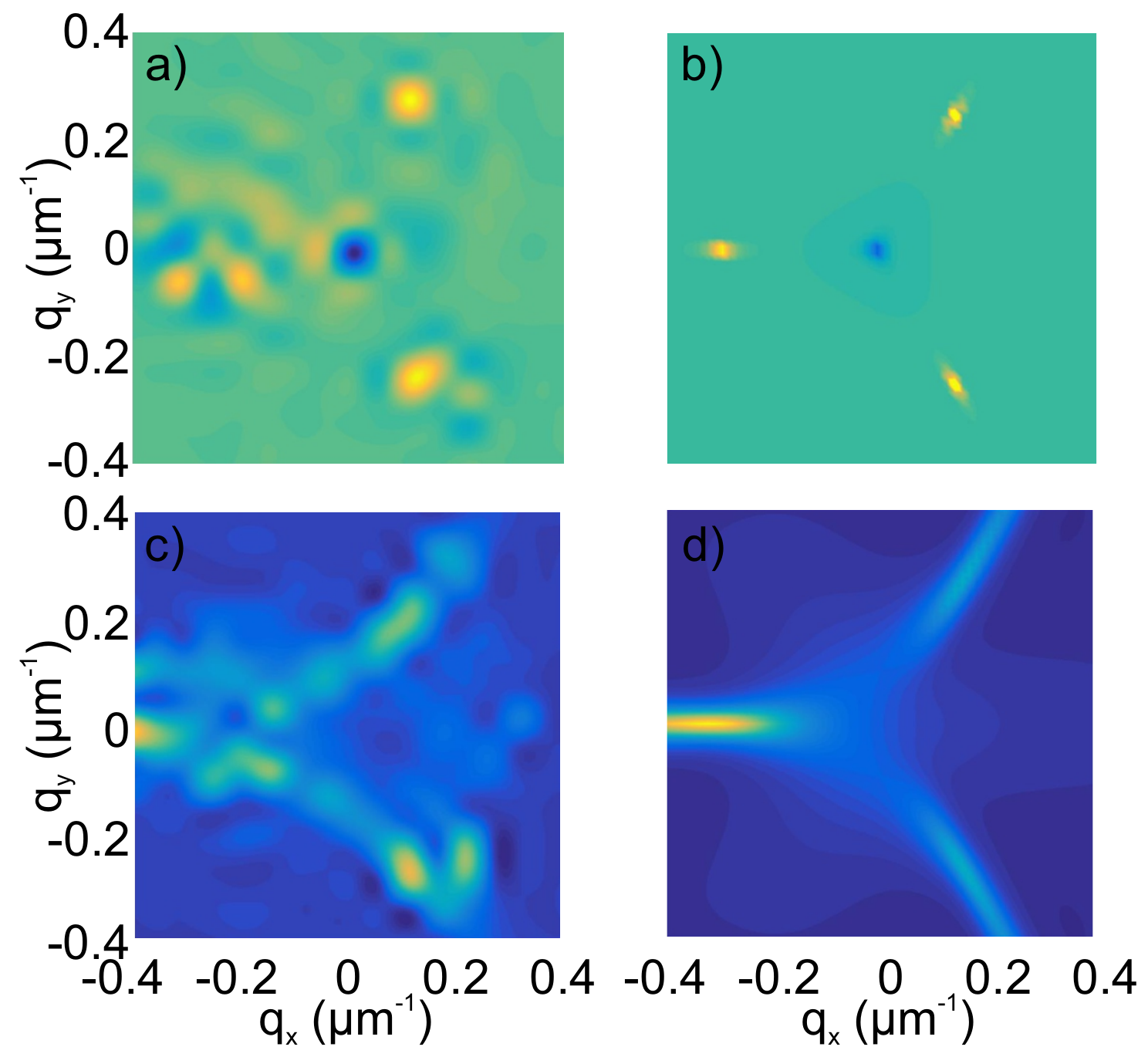}
 \caption{Berry curvature (a) and metric component $g_{xx}$ (c) extracted from numerical simulations based on \eqref{schro} using \eqref{gextract},\eqref{Bextract}. (b)-(d) Tight binding results. Zoom around $K$ point. }
 \label{Fig11}
  \end{center}
 \end{figure}
On figures \ref{Fig9} and \ref{Fig10}, we plot the QGT components in reciprocal space of the two bands below the bandgap (blue and red lines on Fig.~\ref{Fig8}) computed using eqs. \eqref{gHam}, \eqref{BHam}. One can see that the map of these quantities is slightly more complicated than before due to the coupling between the two pseudospins (sublattice and polarization). Indeed one can observe clear reminiscences of the trigonal warping around the corner of the Brillouin zone. One further remark, for the first band each Brillouin zone corner is linked with one negative contribution to the Berry curvature whereas for the second band each of them is associated with three positive contributions. This allows to visualize why the bandgap Chern number will be $C=C_1+C_2=+2$. However, while the total Chern number remains unchanged as long as the gap does not close, the local Berry curvature can be redistributed between the two bands below the bandgap as a function of the parameters: geometry can be smoothly deformed without changing the overall topology.

In numerical simulations, the main difference with respect to the staggered (but spinless) honeycomb lattice is the necessity to extract the phase difference between the pillars for a single spin component ($\sigma^-$), which can be experimentally realized by making interfere the light emitted by different pillars \cite{Sala2015} using an additional polarizer. After solving the Schr\"odinger equation Eq. \eqref{schro} for a lattice of $26\times 26$ unit cells, taking into account the TE-TM coupling and Zeeman splitting, we have extracted the Berry curvature close to one of the $K$ points of the 2nd band (the one on the left of Fig.~\ref{Fig10}). The results of the extraction are shown in Fig.~\ref{Fig11} (a,c). Zoomed tight-binding results are plotted on Fig.~\ref{Fig11} (b,d) for clarity. The parameters are inherently different from the ones of figures \ref{Fig9} and \ref{Fig10}: $\delta J=0.44$, $\Delta_z=0.1$: the TE-TM splitting has been enhanced to allow clear observation of the trigonal warping. As a consequence of the latter, we observe 3 points with positive Berry curvature and 1 point with negative Berry curvature \cite{Bleu2017} in the middle (QGT components have been redistributed with respect to figures \ref{Fig9} and \ref{Fig10}). The positive point on the left is less visible because it is not on the edge of the first Brillouin zone.

\section{Conclusions}
To conclude, we have presented a method of direct extraction of the quantum geometric tensor components in reciprocal space from the results of the optical measurements. We demonstrate the successful application of this method to two different two-band systems: a planar cavity and a staggered honeycomb lattice. In the second part, we generalize the method to a four-band system, considering a regular honeycomb lattice with TE-TM splitting and Zeeman splitting as an example. The numerical experiment accuracy enables to observe the interesting patterns of the quantum metric and the Berry curvature, as the signature of the trigonal warping in the case of a four-component spinor, which allows to be optimistic for future experiments.

The access to these geometrical quantities will allow to increase our understanding of each of the systems presented in the different examples,  where the QGT could affect the transport phenomena (e.g. via the anomalous Hall effect). The knowledge of the geometry of the quantum space is of a fundamental general interest by itself. Finally, a similar method could be applied to get the information on the symmetry of the underlying lattice of the Universe in various lattice models \cite{Beane2014}.

\begin{acknowledgments}
We acknowledge the support of the project "Quantum Fluids of Light"  (ANR-16-CE30-0021), of the ANR Labex Ganex (ANR-11-LABX-0014), and of the ANR Labex IMobS3 (ANR-10-LABX-16-01). D.D.S. acknowledges the support of IUF (Institut Universitaire de France). We thank A. Amo, A. Bramati, J. Bloch, M. Milicevic, and S. Koniakhin for useful discussions.
\end{acknowledgments}

\appendix
\section{Generality of the bispinor wavefunction}
A bispinor is composed of 4 complex numbers, which we write here in the polar form:
\[\left| \psi  \right\rangle  = \left( {\begin{array}{*{20}{c}}
{{c_1}{e^{i{\varphi _1}}}}\\
{{c_2}{e^{i{\varphi _2}}}}\\
{{c_3}{e^{i{\varphi _3}}}}\\
{{c_4}{e^{i{\varphi _4}}}}
\end{array}} \right)\]
with a normalization condition ($c_i$ are real and positive)
\[c_1^2 + c_2^2 + c_3^2 + c_4^2 = 1\]
Let us first deal with the phase of the bispinor:
\begin{widetext}
\[\left( {\begin{array}{*{20}{c}}
{{c_1}{e^{i{\varphi _1}}}}\\
{{c_2}{e^{i{\varphi _2}}}}\\
{{c_3}{e^{i{\varphi _3}}}}\\
{{c_4}{e^{i{\varphi _4}}}}
\end{array}} \right) = {e^{i{\varphi _4}}}\left( {\begin{array}{*{20}{c}}
{{c_1}{e^{i\left( {{\varphi _1} - {\varphi _4}} \right)}}}\\
{{c_2}{e^{i\left( {{\varphi _2} - {\varphi _4}} \right)}}}\\
{{c_3}{e^{i\left( {{\varphi _3} - {\varphi _4}} \right)}}}\\
{{c_4}}
\end{array}} \right) = {e^{i{\varphi _4}}}\left( {\begin{array}{*{20}{c}}
{{c_1}{e^{i\left( {{\varphi _1} - {\varphi _2}} \right)}}{e^{i\left( {{\varphi _2} - {\varphi _4}} \right)}}}\\
{{c_2}{e^{i\left( {{\varphi _2} - {\varphi _4}} \right)}}}\\
{{c_3}{e^{i\left( {{\varphi _3} - {\varphi _4}} \right)}}}\\
{{c_4}}
\end{array}} \right)\]
\end{widetext}
This allows us to group the phase terms as in the main text:
\begin{widetext}
\[\left| \psi  \right\rangle  = {e^{i{\varphi _4}}}\left( {\begin{array}{*{20}{c}}
{{c_1}{e^{i\left( {{\varphi _1} - {\varphi _2}} \right)}}{e^{i\left( {{\varphi _2} - {\varphi _4}} \right)}}}\\
{{c_2}{e^{i\left( {{\varphi _2} - {\varphi _4}} \right)}}}\\
{{c_3}{e^{i\left( {{\varphi _3} - {\varphi _4}} \right)}}}\\
{{c_4}}
\end{array}} \right) = {e^{i{\varphi _4}}}\left( {\begin{array}{*{20}{c}}
{\left( {\begin{array}{*{20}{c}}
{{c_1}{e^{i\left( {{\varphi _1} - {\varphi _2}} \right)}}}\\
{{c_2}}
\end{array}} \right){e^{i\left( {{\varphi _2} - {\varphi _4}} \right)}}}\\
{\left( {\begin{array}{*{20}{c}}
{{c_3}{e^{i\left( {{\varphi _3} - {\varphi _4}} \right)}}}\\
{{c_4}}
\end{array}} \right)}
\end{array}} \right)\]
\end{widetext}
Now let us deal with the real positive coefficients $c_i$, keeping in mind the normalization condition. We can rewrite the latter as:
\[c_1^2 + c_2^2 + c_3^2 + c_4^2 = {\left( {\sqrt {c_1^2 + c_2^2} } \right)^2} + {\left( {\sqrt {c_3^2 + c_4^2} } \right)^2} = 1\]
To simplify the derivation, let us define new variables $a$ and $b$ as:
\[a = \sqrt {c_1^2 + c_2^2} \]
and
\[b = \sqrt {c_3^2 + c_4^2} \]
The normalization condition then reads:
\[{a^2} + {b^2} = 1\]
For any possible values of $a$ and $b$ which satisfy this equation, there exists an angle $\xi$ such that $a=\cos\xi$ and $b=\sin\xi$. This angle can be obtained as $\xi=\arctan b/a$. In our calculations in the main text, we are rather using $\theta_{AB}=2\xi$, which means that $a=\cos\theta_{AB}/2$ and $b=\sin\theta_{AB}/2$. Since $\xi$ exists, $\theta_{AB}$ exists as well. Let us now rewrite the amplitudes of the bispinor as follows:
\[\left( {\begin{array}{*{20}{c}}
{{c_1}}\\
{{c_2}}\\
{{c_3}}\\
{{c_4}}
\end{array}} \right) = \left( {\begin{array}{*{20}{c}}
{\frac{{{c_1}}}{{\cos \xi }}\cos \xi }\\
{\frac{{{c_2}}}{{\cos \xi }}\cos \xi }\\
{\frac{{{c_3}}}{{\sin \xi }}\sin \xi }\\
{\frac{{{c_4}}}{{\sin \xi }}\sin \xi }
\end{array}} \right) = \left( {\begin{array}{*{20}{c}}
{\left( {\begin{array}{*{20}{c}}
{\frac{{{c_1}}}{{\cos \xi }}}\\
{\frac{{{c_2}}}{{\cos \xi }}}
\end{array}} \right)\cos \xi }\\
{\left( {\begin{array}{*{20}{c}}
{\frac{{{c_3}}}{{\sin \xi }}}\\
{\frac{{{c_4}}}{{\sin \xi }}}
\end{array}} \right)\sin \xi }
\end{array}} \right)\]
Then, we can see that for the upper part of the bispinor, the following expression is verified (based on the definition of $a=\cos\xi$ above):
\[{\left( {\frac{{{c_1}}}{{\cos \xi }}} \right)^2} + {\left( {\frac{{{c_2}}}{{\cos \xi }}} \right)^2} = \frac{{c_1^2 + c_2^2}}{{{{\cos }^2}\xi }} = \frac{{c_1^2 + c_2^2}}{{c_1^2 + c_2^2}} = 1\]
The two coefficients in the parenthesis in the upper part of the bispinor are therefore normalized to 1, and we can apply the same reasoning to them: there exists an angle $\xi_A$ such that $c_1/\cos\xi=\cos\xi_A$ and $c_2/\cos\xi=\sin\xi_A$ (this angle is given by $\xi_A=\arctan c_2/c_1$). Again, in the main text we have used a twice larger angle $\theta_A=2\xi_A$. Similar reasoning applies as well to the lower part of the bispinor, which allows to find $\theta_B=2\arctan c_4/c_3$.

We have thus demonstrated that an arbitrary bispinor can be written in the form given in the main text:
\[\left| \psi  \right\rangle  = \left( {\begin{array}{*{20}{c}}
{{c_1}{e^{i{\varphi _1}}}}\\
{{c_2}{e^{i{\varphi _2}}}}\\
{{c_3}{e^{i{\varphi _3}}}}\\
{{c_4}{e^{i{\varphi _4}}}}
\end{array}} \right) = {e^{i{\varphi _4}}}\left( {\begin{array}{*{20}{c}}
{\left( {\begin{array}{*{20}{c}}
{\cos {\xi _A}{e^{i\left( {{\varphi _1} - {\varphi _2}} \right)}}}\\
{\sin {\xi _A}}
\end{array}} \right)\cos \xi {e^{i\left( {{\varphi _2} - {\varphi _4}} \right)}}}\\
{\left( {\begin{array}{*{20}{c}}
{\cos {\xi _B}{e^{i\left( {{\varphi _3} - {\varphi _4}} \right)}}}\\
{\sin {\xi _B}}
\end{array}} \right)\sin \xi }
\end{array}} \right)\]

 \bibliography{biblio} 
\end{document}